\documentclass[12pt]{iopart}
\usepackage{graphicx}
\newcommand{\ket}[1]{| \, #1 \rangle}

\begin{document}

\title{An optical fusion gate for W-states}

\author{\c{ S} K \"Ozdemir$^{1,2}$, E Matsunaga$^2$, T Tashima$^2$, T Yamamoto$^2$, \\ M Koashi$^2$ and N Imoto$^2$}

\address{$^1$Dept. of Electrical and Systems Engineering, Washington University, St. Louis, MO 63130, USA}
\address{$^2$Graduate School of Engineering Science, Osaka University, Toyonaka, Osaka 560-8531, Japan}
\ead{ozdemir@ese.wustl.edu}
\begin{abstract}
We introduce a simple optical gate to fuse arbitrary size polarization
entangled W-states to prepare larger W-states. The gate requires a polarizing beam splitter (PBS), a half wave plate
(HWP) and two photon detectors. We study numerically and analytically the necessary resource consumption for preparing larger W-states by fusing smaller ones with the proposed fusion gate. We show analytically that resource requirement scales at most sub-exponentially with the increasing size of the state to be prepared. We numerically determine the resource cost for fusion without recycling where W-states of arbitrary size can be optimally prepared. Moreover, we introduce another strategy which is based on recycling and outperforms the optimal strategy for non-recycling case.
\end{abstract}

\maketitle

\section{Introduction}

Quantum entanglement is at the heart of many
quantum information processing (QIP) tasks such as quantum
teleportation \cite{s0}, quantum key distribution (QKD) \cite{s1} and quantum computation\cite{s2}. Among the multipartite entangled
states, W, GHZ and cluster states form inequivalent classes in
that they cannot be transformed into each other by local
operations and classical communication (LOCC) \cite{s3}. Recent
studies have shown that for specific type of QIP tasks, specially
designed multi-partite entangled states are required: Cluster
states have been proposed as a universal substrate for measurement
based quantum computation \cite{s10.1}, GHZ class has not only
been shown to be useful for quantum teleportation\cite{s4},
quantum secret sharing \cite{s5,s6} and QKD\cite{s7} but also to
be the only one for reaching consensus in distributed networks
when no classical post-processing is allowed\cite{s8}. On the
other hand, W-class is proposed as a resource for QKD\cite{s9} and
for the optimal universal quantum cloning machine\cite{s10} as
well as shown to be the only pure state to exactly solve the
problem of leader election in anonymous quantum networks
\cite{s8}. It is thus important to efficiently prepare states of
different classes not only for practical applications but also for
the fundamental study of quantum information.

It is known that starting with Bell pairs, probabilistic quantum
parity checking gates, the so-called {\it fusion gates}, can be
efficiently used to grow large scale cluster states
\cite{s22.1}. A GHZ state can be prepared and expanded by one photon with each successful application of the same fusion gate \cite{s22.1,s12}. So far a similar study on n-partite W-states,
defined as $\ket {W_n}=\ket{n-1,1}/\sqrt{n}$ where $\ket{n-k,k}$
is the sum of all states with $n-k$ zeros ($H$-polarized photon) and
$k$ ones ($V$-polarized photons),  have not been carried out.
Given that W states are optimal in the amount of pairwise
entanglement when $n-2$ parties are discarded \cite{s3,s3.6} and
have persistency of $n-1$ which is much larger than those of GHZ
and cluster states \cite{s4.6}, it becomes a necessity to probe the bounds on the
efficiency of preparing large scale W-states and the resource
requirements. It is also worth noting that $\ket {W_n}$ constitute
an entangled web and fully interconnected quantum network. Therefore, such
a study may shed light on the scalability of entangled webs of
W-states \cite{s4.6.1}.

There have been many proposals on preparation and manipulation of W-states and their experimental implementations in photons
\cite{s11,s14,s15,s16,s20,s3.4,s21,s22,tas1}, trapped ions \cite{s16.2,s16.3,s27}, and NMR systems \cite{s16.4}. Recently, we demonstrated experimentally that a three-partite W-state can be prepared from two Bell pairs using LOCC \cite{tas1}. Moreover, we proposed
two schemes for the preparation and expansion of N-partite W-states \cite{tas2,tas3}.

In Ref. \cite{tas2}, we proposed an optical gate formed by a pair of 50:50 beamsplitters which accepts one photon from $\ket {W_n}$ to expand it into $\ket {W_{n+2}}$ with a success probability of $(n+2)/16n$ using an ancillary state of two H-polarized photons $\ket{2_H}$. The gate can be cascaded in which each successful gate operation increases the size of W-state by two photons: Cascading this gate $k$-times prepares the state $W_{n+2k}$ with a success probability of $2^{-4k}(1+2k/n)$. We experimentally demonstrated this optical gate and successfully generated three-photon and four-photon polarization-based W-states \cite{tas2010}. In Refs. \cite{tas3,Gong}, on the other hand, it was shown that $\ket {W_n}$ can be expanded to $\ket {W_{n+1}}$ with a success probability of $(n+1)/5n$ using a polarization dependent beamsplitter and an ancillary state of H-polarized single photon $\ket{1_H}$. The cascade operation of this gate expands a W-state by one photon at each successful step: Cascaded application of this gate $k$-times prepare the state $W_{n+k}$ with a success probability of $5^{-k}(1+k/n)$. As it is clear, for both of these gates the success probability decreases exponentially as $k$ increases. Recycling in case of failure is not possible; thus resource required to prepare a large W-state scales exponentially.

In this paper we present a theoretical proposal for preparing large scale W-state networks using a fusion mechanism. We have previously introduced the basic principles of this W-state fusion mechanism and the gate for its realization in Refs. \cite{SahQCMC,SahBrasil,TashimaPhDthesis}. Here, we compare resource requirements for preparing arbitrarily large W-states using this fusion mechanism under various scenarios. We show that
resource for the proposed fusion mechanism scales subexponentially. Although we
phrase our proposal for qubits encoded in horizontal (H) and
vertical (V) photon polarization, the fusion mechanism is of wider
applicability. The primary resource we will make use of is three
photon polarization entangled W-states $\ket {W_3}$. These can be
prepared using the gate given in Ref. \cite{tas2}, with a success probability of $3/16$ starting with a single photon $\ket{1_V}$ and two
photons in Fock state $\ket {\rm 2_{H}}$, and
the gate in Ref. \cite{tas3}, with a success probabilities $3/10$ with a single photon and a Bell pair. Alternatively, two polarization entangled Bell
states can be used to create this initial W-state as shown in Ref. \cite{tas1}.

\section{Fusion Gate for W-states}
\label{sec:fusion-gate-w}
Let us consider the
following scenario: Two spatially separated administrators, Alice
and Bob, decide to merge their small scale entangled webs $\ket
{W_n}_A$ and $\ket {W_m}_B$ into a larger entangled web $\ket
{W_\gamma}_{A\cup B}$ with the help of a trusted third party
Claire. In order to do this each transmits one qubit of their web
to Claire who acts locally on the received two qubits with the
fusion gate and inform them when the task is successful. The
question is whether such a local manipulation is possible or not,
and if possible then how does the scheme looks like. The
polarization entangled W-states of Alice and Bob are
\begin{eqnarray}\label{N01}
\ket{W_n}_A=\frac{1}{\sqrt{n}}(\ket{(n-1)_H}_a\ket{1_V}_1+\sqrt{n-1}\ket{W_{n-1}}_a\ket{1_H}_1)
\end{eqnarray}
\begin{eqnarray}\label{N01ab}
\ket{W_m}_B=\frac{1}{\sqrt{m}}(\ket{(m-1)_H}_b\ket{1_V}_2+\sqrt{m-1}\ket{W_{m-1}}_b\ket{1_H}_2)
\end{eqnarray} where the photon in the mode $1$ ($2$)
are sent to Claire by Alice (Bob) and those in mode
$a$ ($b$) are kept at Alice's (Bob's) side. We see that the photon pairs
Claire receives are $\ket{1_V}_1\ket{1_V}_2$,
$\ket{1_H}_1\ket{1_V}_2$, $\ket{1_V}_1\ket{1_H}_2$, and
$\ket{1_H}_1\ket{1_H}_2$ with the respective probabilities $P_{\rm
VV}=1/nm$, $P_{\rm HV}=(n-1)/nm$, $P_{\rm VH}=(m-1)/nm$, and
$P_{\rm HH}=(n-1)(m-1)/nm$.

\begin{figure}[h]
\begin{center}
\includegraphics[width=7cm]{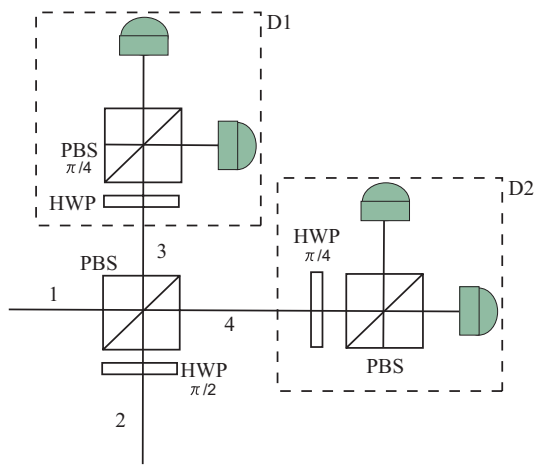} \caption{Non-deterministic fusion gate for W-states: Two spatial modes (1 and 2) with photons
coming from separate W-states are mixed on a PBS, which reflects
vertically (V) polarized photons and transmits horizontally (H)
polarized photons. The half-wave plate (HWP$_{\pi/2}$) at mode 2
transforms $\ket{H}\rightarrow \ket{V}$ and $\ket{V}\rightarrow \ket{H}$. The HWPs $_{\pi/4}$ perform the transformation: $\ket{H}\rightarrow (\ket{H}+\ket{V})/\sqrt{2}$ and $\ket{V}\rightarrow (\ket{H}-\ket{V})/\sqrt{2}$.  The output modes $3$ and $4$ of the first PBS
are measured by the detectors $D1$ and $D2$ which are formed by a HWP, PBS and a pair of photon counters.
\label{fig:1}}
\end{center}
\end{figure}

We start by describing the fusion mechanism which is a parity
check operation (see Fig.1). The photons in two spatial modes are
mixed on a PBS after exchanging the polarization of the photon in
one of the modes by $\pi/2$. After the PBS, photons in the output
modes are measured in $\{\ket{D},\ket{\bar{D}}\}$ basis where $\ket{D}=(\ket{H}+\ket{V})/\sqrt{2}$ and $\ket{\bar{D}}=(\ket{H}-\ket{V})/\sqrt{2}$. The combined action of the HWP and PBS
on the input photons is as follows: $\ket{1_H}_1\ket{1_H}_2\rightarrow
\ket{0}_3\ket{1_H 1_V}_4$, $\ket{1_V}_1\ket{1_V}_2\rightarrow
\ket{1_H 1_V}_3\ket{0}_4$, $\ket{1_H}_1\ket{1_V}_2\rightarrow
\ket{1_H}_3\ket{1_H}_4$, and $\ket{1_V}_1\ket{1_H}_2\rightarrow
\ket{1_V}_3\ket{1_V}_4$ where the subscript numbers denote the
spatial modes of the fusion gate. It is clear that a coincidence detection between the detectors D1 and D2 takes place when the photons in modes 1 and 2 have orthogonal polarizations, and no coincidence is observed when the photons in modes 1 and 2 have the same polarizations.
Moreover, the detectors cannot discriminate between the two cases which lead to coincidence detection.

The case when D1 detects photons but not D2 implies that both of the initial W-states have lost their V-polarized photons. Therefore, the remaining photons will all be H-polarized. Thus, we end up with a product state (networks are destroyed). Such events which we call as {\it failure} take place with a probability of $P_{\rm f}(W_n,W_m)=P_{\rm VV}=1/nm$. On the other hand, for the case where D2 detects photons but D1 does not implies that each of the initial W-states have lost one H-polarized photon. Thus, we will have two separate W-states with smaller number of qubits, $\ket{W_{n-1}}$ and $\ket{W_{m-1}}$ with probability $P_{\rm r}(W_n,W_m)=P_{\rm HH}=(n-1)(m-1)/nm$. This shortened W-states can be recycled using the same fusion mechanism later.

Now let us look at the cases which lead to a coincidence detection closely. When both D1 and D2 detect photons in the same state $\ket{D}$ (or $\ket{\bar{D}}$), the state of the remaining photons becomes
\begin{eqnarray}\label{N01a}
\hspace{-10mm}&&\frac{1}{2\sqrt{nm}}\left(\sqrt{n-1}\ket{W_{n-1}}_a\ket{(m-1)_H}_b+\sqrt{m-1}\ket{(n-1)_H}_a\ket{W_{m-1}}_b\right)\nonumber \\
&&=\frac{1}{2}\sqrt{\frac{n+m-2}{nm}}\ket{W_{n+m-2}}
\end{eqnarray}
where we have used $\sqrt{k}\ket{W_k}=\sqrt{i}\ket{W_i}\ket{(k-i)_H}+\sqrt{k-i}\ket{i_H}\ket{W_{k-i}}$. When one of the detectors detects a photon in $\ket{D}$ and the other in $\ket{\bar{D}}$, the state of the remaining photons will be the same as Eq.(\ref{N01a}) but with a minus sign which can be corrected by applying a $\pi$-phase shift in one of the modes. Thus a coincidence detection signals the successful fusion operation and the preparation of a W-state with $n+m-2$ photons with the success probability $P_{\rm s}(W_n,W_m)=(n+m-2)/nm$. Note that an attempt to fuse $\ket{W_2}$ with $\ket{W_n}$ will not expand the W-state; successful events will prepare only the state $\ket{W_n}$. Expansion requires that both $n$ and $m$ are greater than or equal to 3.

\begin{figure}
\begin{center}
\includegraphics[width=10cm]{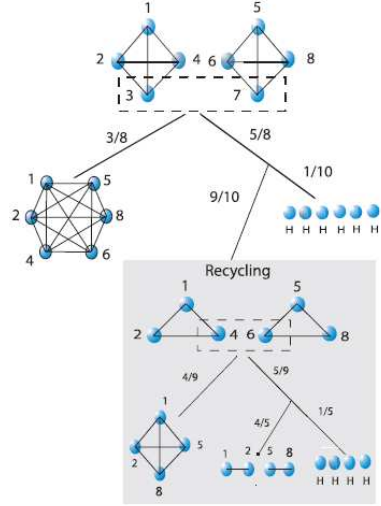} \caption{An example of fusing two W-states. Fusion gate operates
on the qubits in the dotted box. Dark box indicate the recycling process.\label{fig:3}}
\end{center}
\end{figure}

We give an example of fusion operation in Fig.\ref{fig:3}, which show the success, failure or recycling processes. Recycling is performed until either a Bell pair $\ket{W_2}=(\ket{HV}+\ket{VH})/\sqrt{2}$ or a
product state is obtained.. In principle, the resultant Bell states can be further recycled to prepare a $\ket{W_3}$ using the gate introduced in Ref. \cite{tas1}; however, in this study we do not consider such recycling.

As the above discussions show the fusion gate for W-states differ from that for Cluster states in two ways: (i) Fusion gates for cluster states operate with constant success probability of $P_s$ regardless of the size of the cluster states attempted to be fused. However, success probability of fusion gate for W-state depends on the size of the W-states: As $n$ and $m$ increases, $P_f$ decreases but this does not necessarily leads to an increase of the same order in $P_s$, instead the probability $P_r$ of recyclable events increases. (ii) Failure events in cluster state fusion leads to two cluster states shortened by one qubit, similarly to the recyclable events in fusion of W-states. On the other hand, a failure in fusion for W-states leads to complete destruction of both W-states. These make the analysis of fusion and expansion of W-states much more difficult.

\section{Cost of Preparing Arbitrary-Size W states} In this section, we
compare the performances of various strategies using the proposed fusion gate for the preparation of arbitrary size W-states. We will answer the question ``How does the required resource to prepare a W-state of N-photons $\ket{W_N}$ scale?" in various scenarios with and without recycling process.

In this section, we switch to a new index to represent the size of W
states, which differs from the orignal by 2:
\begin{equation}
 \ket{w_n}\equiv \ket{W_{n+2}}.
\end{equation}
The benefit of the lower-case notation is to make the result of successful
fusion more intuitive: successful fusion of $w_m$ and $w_n$ simply
produces $w_{m+n}$. The recyclable outcome leaves the states in the same
form as in the original notation, namely, $w_{m-1}$ and $w_{n-1}$.
The probabilities associated with the three outcomes are now written as
\begin{eqnarray}
 P_{\rm s}(w_n,w_m) &=& (n+m+2)/[(n+2)(m+2)]\\
 P_{\rm r}(w_n,w_m) &=& (n+1)(m+1)/[(n+2)(m+2)]\\
 P_{\rm f}(w_n,w_m) &=& 1/[(n+2)(m+2)]
\end{eqnarray}

\begin{figure}[h]
\begin{center}
\includegraphics[width=5cm]{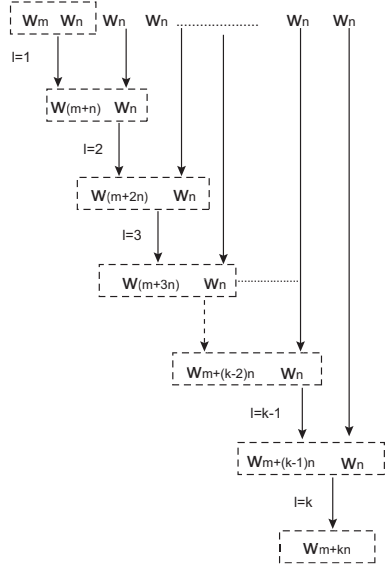}\caption{Linear growth of W-states using the proposed fusion gate. Consecutive successful events are shown and recycling is not performed. Dashed boxes denote the fusion operation. Note that actual sizes of the W-states are found by adding $2$ to the sizes given in the figure.
\label{fig:3a1}}
\end{center}
\end{figure}

We use the notation $R[w_m]$ for the resource cost (i.e., the number of $\ket{w_1}$ states required) of producing state
$w_m$, which is a $(m+2)$ qubit W state $\ket{W_{m+2}}$.
In our analysis, we consider $\ket{W_3}\equiv\ket{w_1}$ as the basic
resource provided with unit cost, i.e., $R[w_1]=R[W_3]=1$.
For $m\ge 2$, the value of $R[w_m]$ will vary depending on the strategies.

When we do not use recycling and try to produce $w_{m+n}$ from $w_n$ and
$w_m$, the costs are simply related as
\begin{eqnarray}\label{N05}
R[w_{n+m}]&=&\frac{1}{P_{\rm s}(w_n,w_m)}(R[w_{n}]+R[w_{m}])\nonumber\\
&=&\frac{(m+2)(n+2)}{n+m+2}(R[w_{n}]+R[w_{m}]),
\end{eqnarray}
which is frequently used in the later analysis. The cases with recycling
are more complicated and will be treated separately below.

In the following, we treat this problem for linear and exponential growth strategies with and without recycling and derive analytical bounds for resource complexity. We provide the optimal strategy for fusion without recycling, and also introduce a strategy based on fusing $W$ states of similar sizes which turns out to provide the best resource scaling among all the strategies considered here.

\subsection{Linear Growth Strategies}
Here, we analyze the resource requirements for linear growth strategies with and without recycling. The strategy is based on repeated fusion of a fixed-size W-state to an already existing W-state (see Fig.\ref{fig:3a1}). Let us assume that we want to expand $w_m$ by fusing it with $w_n$ repeatedly. If fusing is successful, bring another $w_n$ and fuse it with the state prepared in the previous level. In this way after the $k$-th level of successively successful gate operations, the state $w_{m+kn}$ is prepared with the probability $(m+kn+2)/(m+2)(n+2)^{k}$. At each level of successful operation, the size of the state increases by $n$, i.e., $\{m,\underline{n}\}\rightarrow\{m+n,\underline{n}\}\rightarrow\{m+2n,\underline{n}\}\ldots \rightarrow\{m+(k-1)n,\underline{n}\}\rightarrow\{m+kn\}$.
\subsubsection{Linear growth without recycling}
Resource required for expanding the state $w_m$ by $n$ is given in Eq. \ref{N05} as $R[w_{m+n}]=P_{\rm s}^{-1}(w_m,w_n)(R[w_{m}]+R[w_{n}])$ which can be re-written as
\begin{eqnarray}\label{N07777}
(m+n+2)R[w_{m+n}]=(m+2)(n+2)(R[w_{m}]+R[w_{n}]).
\end{eqnarray} Similarly, the cost $R[w_{m+(k+1)n}]$ of
preparing the state by fusing $w_n$ with the state $w_{m+nk}$ prepared
in a previous fusion operation becomes
\begin{eqnarray}\label{N07779}
\hspace{-15mm}(m+(k+1)n+2)R[w_{m+(k+1)n}]=(m+kn+2)(n+2)(R[w_{m+kn}]+R[w_{n}]).
\end{eqnarray} Defining $r_{k}=(m+kn+2)R[w_{m+kn}]$ and $\xi=(n+2)R[w_{n}]$, we can re-write Eq. \ref{N07779} as
\begin{eqnarray}\label{N07780a}
r_{k+1}=(n+2)r_k+\xi(m+kn+2)
\end{eqnarray} which is a first order inhomogenous recurrence equation. Solving Eq.(\ref{N07780a}) yields
\begin{eqnarray}\label{N07780}
r_{k}=(r_0-\beta)(n+2)^k+\alpha k +\beta
\end{eqnarray}
where $\alpha=-\xi n/(n+1)$, $\beta=(\alpha-\xi(m+2))/(n+1)$, and $r_0=(m+2)R[m]$. Subsequently, we can find the cost $R[w_{m+kn}]$ of preparing the state $w_{m+kn}$ as $R[w_{m+kn}]=r_k/(m+kn+2)$. Here we consider the case $m=n=1$ where $\ket {w_1}= \ket {\rm W_3}$ is
repeatedly fused. Then we have $R[w_{k+1}]=r_k/(k+3)$ where $r_{k}$ is
found by setting $n=1$ and $m=1$ in Eq. (\ref{N07780}). Consequently, we can find the cost $R[w_{N}]$ of preparing the state $w_{N}$ by setting $k\rightarrow N-1$ as
\begin{eqnarray}
R[w_N]&=&\frac{1}{N+2}\left(\frac{11}{4}3^{N}-\frac{3}{2}N-\frac{15}{4}\right)
\end{eqnarray} which is depicted in Fig. \ref{fig:5b2} (red colored $+$). Thus we conclude that the cost of preparing the state $w_N$ by {\it linear growth} strategy is of the order $O(3^N)$, that is it increases exponentially with $N$.
\subsubsection{Linear growth with recycling} As shown in
Sec.~\ref{sec:fusion-gate-w}, recyclable failure of the fusion gate,
which takes place with probability $P_{\rm
r}(w_n,w_m)=(n+1)(m+1)/(n+2)(m+2)$,  leads to a reduction in the size of
the initial W states by one qubit, e.g.,
$\{m,n\}\rightarrow\{m-1,n-1\}$. Therefore, the remaining W states can
be recycled. Here we include this recycling into the linear-growth
strategy and see how the cost $R[w_m]$ changes.

The averaged cost $R[w_{m+1}]$ can be written as
$R[w_{m+1}]=R[w_{m}]+\Delta_{m+1}$, where $\Delta_{m+1}$ is the averaged cost of
creating $w_{m+1}$ when a shorter W state $w_{m}$ is given.
Let us calculate $\Delta_{m+1}$ as follows.
Suppose that state $w_{m}$ is given, and we apply a fusion gate
to it with $w_{1}$, by paying a unit cost.
At probability $p_m\equiv P_{\rm s}(w_m,w_1)$, the gate succeeds,
and no additional cost is required.
At probability $q_m\equiv P_{\rm r}(w_m,w_1)$,
we are left with state $w_{m-1}$, and it takes additional cost of
$\Delta_{m}+\Delta_{m+1}$ to obtain $w_{m+1}$.
Finally, at probability $1-p_m-q_m$, the gate fails completely and
we need full cost $R[w_{m+1}]$ to produce $w_{m+1}$.
These observations lead to
\begin{equation}
 \Delta_{m+1}=1+q_m(\Delta_{m}+\Delta_{m+1})+(1-p_m-q_m)R[w_{m+1}]
\nonumber
\end{equation}
and thus we have
\begin{equation}
p_m R[w_{m+1}]= R[w_{m}]+1-q_m R[w_{m-1}].
\end{equation}
This recursive formula can be numerically solved with
$R[w_{1}]=1$ and $R[w_{2}]=9/2$, which is
depicted in Fig. \ref{fig:5b2} (red colored dotted box).
When $m$ is large, $p_m\sim 1/3$ and $q_m\sim 2/3$
lead to $R[w_{m+1}]-R[w_{m}]\sim 2(R[w_{m}]-R[w_{m-1}])$.
We thus conclude that although
recycling reduces the required resources from $O(3^m)$ to $O(2^m)$,
it does not change the resource scaling law: Regardless of whether
recycling is performed or not, required amount of resource to prepare a
desired state using linear-growth strategies scales exponentially.

\subsection{Optimal Strategy without Recycling}

We have seen that strategies to grow W states by a constant amount
at each step is not so efficient, even if recycling is introduced.
We should thus turn to other strategies for efficiency.
In this subsection, we numerically determine the optimal cost
over all the strategies without recycling.

Let us start from simple examples. In order to produce $w_2$, the only
way is to fuse two $w_1$ states, since $w_2$ cannot be produced when
larger W states are fused under the assumption of no recyclinig.
The optimal cost $R[w_2]_{\rm opt}$ is thus given by
$P_{\rm s}(w_1,w_1)^{-1}(1+1)=9/2$.
Similarly, there is only one way to produce $w_3$,
leading to
$R[w_3]_{\rm opt}=P_{\rm s}(w_2,w_1)^{-1}(1+9/2)=66/5$.
In the case of $w_4$, on the other hand,
there are two possible ways, $\{w_1,w_3\}$ and $\{w_2,w_2\}$ with the
respective successful fusion probabilities of $2/5$ and $3/8$. Since
we know the optimal costs for preparing $w_2$ and $w_3$, we calculate
the cost of preparing $w_4$ from $\{w_1,w_3\}$ as
$(5/2)(1+66/5)=71/2$, whereas that from $\{w_2,w_2\}$ as
$(8/3)(9/2+9/2)=24$. The latter strategy is better, and hence
$R[w_4]_{\rm opt}=24$. In this way, the optimal cost of any state
can be numerically calculated using the recursive formula
\begin{equation}\label{N2345a}
R[w_m]_{\rm opt}=\min_{k=1,\ldots,m-1}\frac{R[w_{k}]_{\rm opt}+R[w_{m-k}]_{\rm opt}}{P_{\rm s}(w_{k},w_{m-k})}.
\end{equation}
The calculated values of $\{R[w_N]_{\rm opt}\}$ are
presented in Fig.\ref{fig:5b2} (green colored dotted circle), which suggests a sub-exponential
resource scaling.

\begin{figure}
\begin{center}
\includegraphics[width=7.5cm]{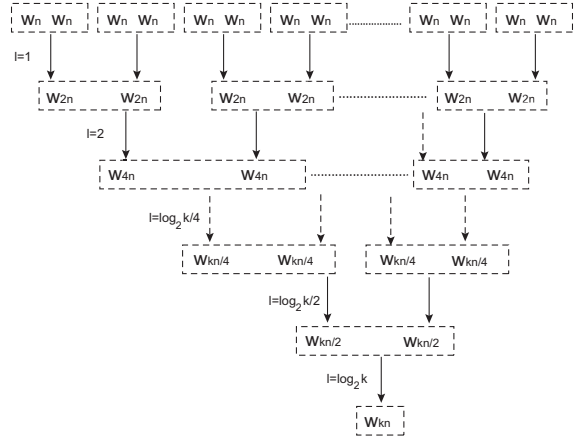}\caption{Exponential growth of W-states using the proposed fusion gate. W-states of the same size are fused. Consecutive successful events are shown and recycling is not performed. Dashed boxes denote the fusion operation. Note that actual sizes of the W-states are found by adding $2$ to the sizes given in the figure.
\label{fig:3b1}}
\end{center}
\end{figure}

\subsection{Exponential Growth Strategy without Recycling}

Although the discussion in the previous subsection enables us to
numerically calculate the optimal cost in the case of no recycling,
it does not tell us how the optimal strategy look like or how the
cost scales in the limit of large target size.
Here we consider a specific strategy without recycling,
based on fusing two states of the same size to double it.
We show that this strategy works under the optimal cost,
and derive an analytical expression of the cost in order to see the
scaling over the target size.

Here we only consider production of a state $w_N$ whose size is
written as $N=2^k$. In what we call an
{\it exponential-growth} strategy, the state $w_N$ is produced by
fusion of two W states $w_{N/2}$ with equal size. The state
$w_{N/2}$ is in turn generated from fusion of state $w_{N/4}$.
When no recycling is done, Eq.~(\ref{N05}) leads to a simple
relation among the costs in this strategy:
\begin{equation}
\label{N010}
R[w_{2^{l+1}}]=\frac{(2^l+2)^2}{2^l+1}R[w_{2^l}].
\end{equation}
If we define $a_l\equiv (1+2^{l-1})R[w_{2^l}]$, we have
$a_{l+1}=2^{l+1}(1+2^{1-l})a_l$ and thus
\begin{eqnarray}\label{N11}
R[w_N]&=&R[w_{2^k}]=\frac{\gamma_k}{1+2^{k-1}}2^{k(k+1)/2}\\
\gamma_k&\equiv& \frac{3}{2}\prod_{l=0}^{k-1}(1+2^{l-1}),
\end{eqnarray}
which is plotted in Fig.\ref{fig:5b2}. We see that the cost coincides
with the optimal cost for no recycling derived in the previous section,
indicating that the strategy of fusing two state of the same size
is very cost-effective.
Since the coefficient $\gamma_k$ is finite, namely,
$\gamma_k \le \lim_{k\to \infty} \gamma_k=21.458\ldots$,
the scaling of the cost in the limit of large $k$ is
$O(2^{k(k+1)/2})$, or equivalently,
\begin{equation}
\label{N16}
 R[w_N]=O(\sqrt{N}N^{\log_2 N /2})
\end{equation}
in the limit of large $N$, which is sub-exponential in $N$.

\subsection{Fusing States of Similar Sizes with Recycling}
We have seen that fusing two W states of the same
size is advantageous in the case of no recycling. Here we propose a strategy with
recycling, which tries to fuse W states of similar sizes. The
performance of the strategy is then evaluated through  Monte Carlo simulations.

Let us classify generated W states into sets $\{S_l\}$ according to their sizes,
such that state $w_m$ belongs to set $S_l$ when $m\in (2^{l-1},2^l]$.
The idea is to perform fusion operation between two states belonging to
the same set.
It is easy to see that fusion of two states $w_m$ and $w_n$ belonging to
the same set $S_l$ will produce state $w_{m+n}$ in $S_{l+1}$ upon
successful operation. If the operation is complete failure, the states
$w_n$ and $w_m$ are discarded.
 In case of recyclable outcome,
the resultant states $w_{m-1}$ and $w_{n-1}$ belong to either
$S_l$ or $S_{l-1}$.

More precisely, our strategy to prepare a state belonging to set $S_{k+1}$ is described
in the following algorithm, which dictates the order in which the
recycled W states should be used. In the description below, $\mu_l$ represents the number
(0,1, or 2) of states already generated in set $S_l$, $R$ is the cost
(the number of consumed states $w_1$), and $\xi$ is a pointer to the
`current' working set.
\begin{enumerate}

\item [(1)] Initially, let all the sets $S_0,\ldots,S_{k+1}$ be empty, $\xi=0$, $R=0$,
 and  $\mu_0,\ldots,\mu_{k+1}=0$.

\item [(2)] Depending on the values of $\xi$ and $\mu_\xi$, do one of the
      following procedure.

($\mu_\xi \le 1$ and $\xi= 0$): Add $w_1$ to $S_0$, $\mu_0\to \mu_0+1$,
$R\to R+1$, and repeat step 2.

($\mu_\xi \le 1$ and $\xi\ge 1$): $\xi\to \xi-1$ (decrement $\xi$) and
      repeat step 2.

($\mu_\xi=2$): Proceed to step 3.

\item [(3)] $S_\xi$ should have two states, which we denote $w_n$ and $w_m$,
      and we apply a fusion gate on them. Empty $S_\xi$ and set
      $\mu_\xi\to 0$.
     Do one of the following procedures depending on the result of the gate operation.

(Complete failure): Go back to step 2.

(Recyclable): Add $w_{n-1}$ and $w_{m-1}$ to appropriate sets ($S_\xi$
      or $S_{\xi-1}$), and update $\mu_\xi$ and $\mu_{\xi-1}$
      accordingly.  Go back to step 2.

(Success): If $\xi=k$, the goal has been achieved with cost $R$ and the procedure ends
      here. Otherwise, add $w_{n+m}$ to $S_{\xi+1}$, $\mu_{\xi+1}\to \mu_{\xi+1}+1$,
      $\xi\to \xi+1$, and go back to step 2.
\end{enumerate}

\begin{figure}[h]
\begin{center}
\includegraphics[width=9cm]{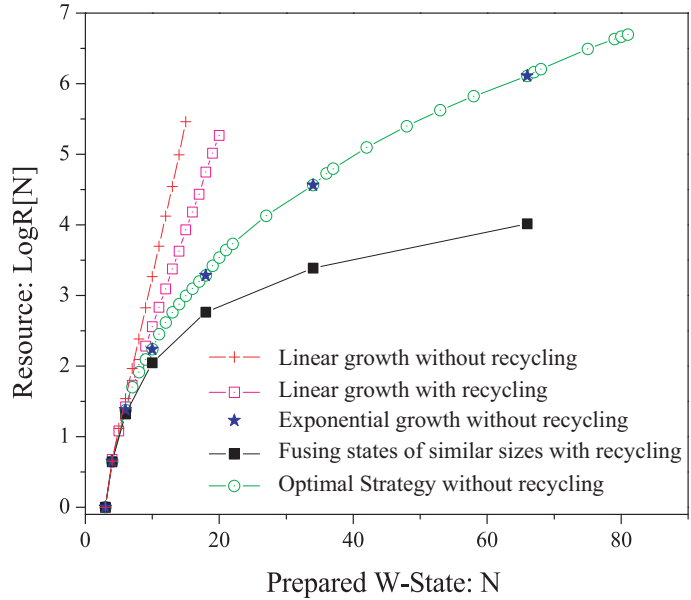}\caption{Comparison of the expected amount of resources R[N] to prepare a $W_N$ for strategies introduced in the text: Linear growth by one, i.e., $\{n,1\}\rightarrow \{n+1\}$, with and without recycling, Exponential growth, i.e., $\{n,n\}\rightarrow\{2n\}$ , without recycling, Fusing W states of similar sizes, and optimal strategy without recycling. Each point corresponds to the average of 1000 trials. Note that sizes given here are the actual sizes of the states.
\label{fig:5b2}}
\end{center}
\end{figure}

The above strategy creates a W-state of (actual) size $N=2^{k}+3$ or larger.
We have done Monte-Carlo simulations for $k=0,\ldots,6$, and
Fig. \ref{fig:5b2} (black square) shows the cost averaged over 1000 runs for each
value of $k$. It is clearly seen that this strategy, based on fusing the states of similar sizes with recycling, outperforms all the other strategies
considered in previous sections.

\section{Conclusion}

In this paper, we introduced an optical fusion gate to fuse W-states to prepare a W-state of larger sizes and discussed the scaling laws for the required resources for preparing a W-state using the proposed fusion gate.

We introduced four different strategies with different resource requirements depending on whether recycling is allowed or not. We first demonstrated both analytically and numerically that resource requirement for linear growth strategies, which consider repeated fusion of a fixed size W-state with the already existing W-state. These strategies scale exponentially regardless of whether re-cycling is performed or not, although recycling allows reduction in the required resources. Then we calculated the optimal cost for fusion without re-cycling. Next, we considered exponential growth strategies in which fusion gate is always applied to two states of the same size. We derived analytical expressions showing that the required resources scale sub-exponentially. Interestingly, non-recycling exponential growth strategy appears to have the same resource scaling as the optimal strategy with no-recycling, implying that the former is the optimal solution for fusion without recycling. Finally, through numerical simulations we demonstrated a strategy, in which states with the closest sizes are fused, provides the best resource scaling for the proposed fusion gate among the strategies investigated in this study.

The proposed fusion gate and discussed fusion strategies outperform the previously proposed W-state preparation and expansion gates in terms of required resources when the size of the state to be prepared is large. Our study does not exclude the possibility of the presence of a better strategy or a strategy with a polynomial scaling for the proposed fusion gate. We hope that this study will initiate further work and continuing discussions on the optimal ways of a fusing/preparing/expanding  W-states and understanding the structure of larger multipartite entangled states.

\ack
This work was supported by Funding Program for World-Leading Innovative R\&D on Science and Technology (FIRST), MEXT Grant-in-Aid for Scientific
Research on Innovative Areas 20104003 and 21102008, JSPS Grant-in-Aid for Scientific Research(C) 20540389, and MEXT Global COE Program. \c{S}.K.\"{O}. thanks Prof.L. Yang of Washington University in St. Louis for her support.

Note added. During the final preparations of this manuscript, we came to know of the work in Ref. \cite{Fujii}. Note that the fusion gate for W-state studied in this manuscript was first proposed by us in Ref. \cite{SahQCMC} and was further detailed in Refs. \cite{SahBrasil} and \cite{TashimaPhDthesis}.

\end{document}